\documentstyle[psfig,12pt]{article}
\def\step{\\[-1.5ex]}

\def\beq{\begin{equation}}
\def\eeq{\end{equation}}

\def\bea{\arraycolsep .1em \begin{eqnarray}}
\def\eea{\end{eqnarray}}
\def\Tr{{\rm Tr}}

\let\de=\delta

\def\eq#1{(\ref{#1})}
\def\Eq#1{Eq.~(\ref{#1})}
\def\Es#1{Eqs.~(\ref{#1})}

\def\s0#1#2{\mbox{\small{$ \frac{#1}{#2} $}}}
\def\0#1#2{\frac{#1}{#2}}

\begin{document}
\begin{center}

\thispagestyle{empty}

{\normalsize\begin{flushright}
CERN-TH-2001-169\\
FAU-TP3-01-05\\[12ex] 
\end{flushright}}

\mbox{\large \bf 
Predictive power of renormalisation group flows: a comparison
} \\[6ex]

{Daniel F. Litim 
\footnote{Daniel.Litim@cern.ch}
$\!\!{}^{,*}$
and 
Jan M.~Pawlowski 
\footnote{jmp@theorie3.physik.uni-erlangen.de}}
$\!\!{}^{,\dagger}$
\\[4ex]
{${}^*${\it 
Theory Division, CERN\\ 
CH-1211 Geneva 23.
}\\[1ex]
${}^\dagger${\it 
Institut f\"ur Theoretische Physik III\\ 
Universit\"at Erlangen, 
D-91054 Erlangen.
}}
\\[10ex]
 
{\small \bf Abstract}\\[2ex]
\begin{minipage}{14cm}{\small 
    We study a proper-time renormalisation group, which is based on an
    operator cut-off regularisation of the one-loop effective action.
    The predictive power of this approach is constrained because the
    flow is not an exact one. We compare it to the Exact
    Renormalisation Group, which is based on a momentum regulator in
    the Wilsonian sense.  In contrast to the former, the latter
    provides an {\it exact} flow. To leading order in a derivative
    expansion, an explicit map from the exact to the proper-time
    renormalisation group is established. The opposite map does not
    exist in general. We discuss various implications of these
    findings, in particular in view of the predictive power of the
    proper-time renormalisation group. As an application, we compute
    critical exponents for $O(N)$-symmetric scalar theories at the
    Wilson-Fisher fixed point in $3d$ from both formalisms.}
\end{minipage}
\end{center}

\newpage
\pagestyle{plain}
\setcounter{page}{1}

\noindent
{\bf 1. Introduction}\\[-1ex]

Renormalisation group methods are an essential ingredient in the study
of non-pertur\-bative problems in continuum and lattice formulations
of quantum field theory (QFT). A well-established renormalisation
group, based on the Wilsonian idea of integrating-out infinitesimal
momentum shells, is known as the Exact Renormalisation Group (ERG)
\cite{Wilson,Polchinski,CW,Ellwanger:1994mw,Morris:1994qb} (for recent
reviews, see Refs.~\cite{Bagnuls:2000ae} for scalar and
Ref.~\cite{Litim:1998nf} for gauge theories). The fully integrated ERG
flow of the effective action covers all momenta.  Thus the endpoint of
an ERG flow is the full quantum effective action. An important
advantage of this formalism is its flexibility, when it comes to
approximations or truncations.  This makes it an interesting tool
when, due to the complexity of the problem at hand, approximations are
unavoidable. For non-perturbative effects at strong coupling or for
large correlation lengths, such an approach is essentially
unavoidable. The fact that this flow is derived from first principles
-- in combination with the control of approximate solutions
\cite{Scheme,Litim:2000ci,Litim:2001up,Litim:2001fd} -- is at the root
of reliable approximations and the predictive power of the
formalism.\step

A somewhat similar renormalisation group has been proposed in
Refs.~\cite{Liao:1996fp} and \cite{Liao:1997nm} for scalar theories
and gauge theories, respectively. It is based on a regularised
Schwinger proper-time representation
\cite{Fock:1937dy,Schwinger:1951nm} of the one-loop effective action
\cite{Oleszczuk:1994st}. A variation with respect to the scale leads
to a flow equation. The final proper-time renormalisation group (PTRG)
equation is obtained by a one-loop improvement. On the technical
level, the PTRG is as flexible as the ERG, and approximations schemes
or expansions used for the latter can immediately be taken over to the
former.\step

As opposed to the ERG, the conceptual understanding of the PTRG is
less well developed. There exists no first principle derivation of the
PTRG flow.  Furthermore, the solution to a PTRG flow is {\it not} the
full quantum effective action \cite{Consistency}.  Hence, the PTRG is
at best an approximation to an exact flow.  Currently, it is not known
{\it what} kind of approximation it represents.  Neither is it known
how to relate systematic approximations of the PTRG to systematic
approximations of the full quantum effective action: at present, the
PTRG lacks predictive power.  \step

Here, we take a practitioners point of view in order to shed some
light on these issues. We provide a partial control for PTRG flows by
comparing them to ERG flows to leading order of the derivative
expansion. In this approximation we discuss the map which connects
PTRG and ERG. We show that the map exists from ERG to PTRG. We also
show that the opposite map does not exist in general, not even to
leading order in the derivative expansion.  Within a restricted set of
PTRG regulator functions, these findings provide the required link
between approximations to the PTRG and approximations to the physical
theory.  We also discuss the content of those PTRG flows, which cannot
be mapped on ERG flows. As an application of these results, we compute
critical exponents for $3d$ $O(N)$-symmetric scalar theories for both
the ERG and the PTRG.  For specific regulators, our results compare
very well with both experiment and results obtained by other methods.
We discuss the dependence of the results on the regularisation, and
the applicability of an optimisation condition
\cite{Litim:2000ci,Litim:2001up,Litim:2001fd} or a minimum sensitivity
condition \cite{Stevenson,Liao:2000sh,Papp:2000he,Litim:2001fd}.
\\[3ex]

\noindent
{\bf 2. Exact renormalisation group}\\[-1ex]

Let us start with a brief discussion of the main conceptual
ingredients of the ERG. The basic idea leading to an exact flow
equation is the step-by-step integrating-out of fluctuations within a
path integral formulation of quantum field theory in a Wilsonian sense
\cite{Wilson}. This can be seen as a continuum version of the earlier
block spin actions introduced by Kadanoff. A path-integral formulation
of these ideas is due to Polchinski \cite{Polchinski}, while
formulations for the effective action have been given in
Refs.~\cite{CW,Ellwanger:1994mw,Morris:1994qb}. For recent reviews,
see Refs.~\cite{Bagnuls:2000ae,Litim:1998nf}. \step

Within the standard ERG, the integrating-out of fluctuations is
achieved by a {\it momentum} regulator $R_k(q^2)$. It modifies the
effective propagator of the fields, and depends on a fiducial
infra-red scale $k$. The ERG flow of the effective action $\Gamma_k$
with respect to $k$ is given by
\beq\label{ERG}
\partial_t\Gamma_k[\phi]= \frac{1}{2}\Tr \frac{1}{
  \Gamma_k^{(2)}[\phi]+R_k}\, \partial_t R_k 
\eeq 
for bosonic fields $\phi$. Here, the trace denotes a sum over all
momenta and indices, $t=\ln k$ and $ \Gamma_k^{(2)}[\phi]$ stands for
the second derivative of $\Gamma_k$ w.r.t.\ the field $\phi$. \step

The function $R_k(q^2)$ has to satisfy some constraints in order to
provide an infra-red regulator for the effective propagator, and to
ensure that the flow \eq{ERG} interpolates precisely between an
initial (classical) action in the UV and the full quantum effective
action in the IR. The necessary conditions on $R_k$ are summarised as
\bea
\label{1}
\lim_{q^2/k^2\to 0}R_k(q^2)&>&0\\
\label{2}
\lim_{k^2/q^2\to 0}R_k(q^2)&=&0\\
\label{3}
\lim_{k\to \Lambda}R_k(q^2)&\to&\infty\,.
\eea
Eq.~\eq{1} guarantees that $R_k$ provides an IR regularisation,
because mass-less modes are effectively cut-off. The second constraint
\eq{2} ensures that the regulator is removed in the IR limit $k\to 0$.
The condition \eq{3} ensures that the correct initial condition is
reached for $\lim_{k\to \Lambda}\Gamma_k=S_\Lambda$, where $\Lambda$
defines the initial (UV) scale. From now on we will use
$\Lambda=\infty$. Then the regulator $R_k$ can be rewritten in terms
of a dimensionless function $r(y)$ as
\beq\label{r}
R_k(q^2)=q^2\,r(q^2/k^2)
\eeq
The constraints \eq{1} -- \eq{3} on $R_k$ are sufficient to guarantee
that the flow \eq{ERG} interpolates between the initial UV action and
the full quantum effective action $\Gamma$ for $k=0$. In addition,
these conditions imply that the insertion $\partial_t R_k$ in \Eq{ERG}
is peaked as a function of momenta about $q^2\approx k^2$. For large
momenta, $\partial_t R_k$ decays rapidly. Thus contributions to the
flow from UV modes are suppressed. For small momenta, $\partial_t R_k$
either diverges, or it approaches a constant limit. This structure
explains why the flow \eq{ERG} integrates-out only a narrow momentum
shell around the scale $k$.\step

For an explicit computation of the IR effective action $\Gamma$ based
on the ERG approach the specification of the field content, the
initial condition $\Gamma_\Lambda$ and the choice of a particular
regulator is required. As soon as it comes to (unavoidable)
approximations it is very important that the integrated flow
approaches the full quantum effective action. Although approximate
solutions to ERG flows may depend spuriously on the IR regulator
\cite{Scheme}, it has been clarified recently that convergence
properties of approximate solutions towards the physical theory are
controlled by the IR regulator, and improved for specific optimised
choices \cite{Litim:2000ci,Litim:2001up,Litim:2001fd} (see also
Ref.~\cite{Liao:2000sh}). This guarantees that any systematic
truncation of the ERG provides an approximation to the full quantum
effective action. In particular, systematic truncations can be
improved to higher order, leading to better approximations of the
physical theory.
\\[3ex]

\noindent
{\bf 3. Proper-time renormalisation group}\\[-1ex]

As opposed to the derivation of an ERG sketched above there is no
first principle derivation of the PTRG. Instead, the PTRG follows as
an one-loop improvement based on a proper-time regularisation of the
one-loop effective action.  The heuristic derivation of the flow
starts with the well-known expression
\beq\label{1-loop} 
\Gamma^{\rm 1-loop}[\phi]= 
S_{\rm cl}[\phi]
+\012\Tr \ln \frac{\de^2 S_{\rm cl}[\phi]}{\delta \phi\, \delta \phi} 
\eeq 
for the one-loop effective action $\Gamma^{\rm 1-loop}[\phi]$. The
trace in \Eq{1-loop} is ill-defined and requires an UV regularisation
and, in the case of massless modes, also an IR regularisation.
Oleszczuk proposed an UV regularisation by means of a Schwinger
proper-time representation of the trace \cite{Oleszczuk:1994st},
\beq\label{RegDef} 
\Gamma^{\rm 1-loop}[\phi;\Lambda]=
S_{\rm cl}-\s012\int\0{ds}{s}f(\Lambda,s)
\Tr\exp\left(-s\,S_{\rm cl}^{(2)}\right)\,.
\eeq 
The regulator function $f(\Lambda,s)$ provides an UV cut-off $\Lambda$
if $\lim_{s\to 0}f(\Lambda,s)=0$. It implies that $\Gamma^{\rm
  1-loop}[\phi;\Lambda]$ depends on the scale $\Lambda$. Sending the
UV scale to $\infty$ should reduce \Eq{RegDef} to the standard
Schwinger proper-time integral. This happens for the boundary
condition $f(\Lambda\to\infty,s)=1$. A new ingredient has then been
added by Liao \cite{Liao:1996fp}, who noticed that \Eq{RegDef} can be
turned into a simple flow equation by also adding an IR scale $k$,
replacing $f(\Lambda,s)\to f_k(\Lambda,s)$. Introducing another scale
parameter turns $\Gamma^{\rm 1-loop}[\phi;\Lambda]$ into a
$k$-dependent functional. A flow equation for some functional
$\Gamma_k$ with respect to the infra-red scale $k$ is given by
\cite{Liao:1996fp}
\beq\label{PTRG} 
\partial_t \Gamma_k[\phi]= -\s012 \int_0^\infty
\0{ds}{s} \left( \partial_t f_k(\Lambda,s)\right)
\Tr\exp\left(-s\Gamma^{(2)}_k\right)\,  
\eeq 
Here, the classical action has been replaced by the scale-dependent
effective action on the right-hand side of \Eq{PTRG}. This is the
philosophy of an one-loop improvement. Note that in \Eq{PTRG} {\it
  only} the explicit scale dependence due to the regulator term is
considered. A total $k$-derivative would require in addition that the
scale dependence of $\Gamma^{(2)}_k$ is taken into account. It is of
course tempting to identify $\Gamma_k$ as implicitly defined in
\Eq{PTRG} with the full scale dependent effective action. However,
from its derivation we only can conclude that it is a RG improved
approximation to the latter. \step

The PT regulator has to satisfy some constraints similar to those
imposed on $R_k$ \cite{Liao:1996fp,Floreanini:1995aj}. We require that
\bea
\label{f1} 
\lim_{s\to\infty}f_{k\neq 0}(\Lambda,s)&=&0\\
\label{f3} 
\lim_{k\to \Lambda}f_{k}(\Lambda,s)&=&0\\
\label{f4} 
\lim_{\Lambda\to\infty}f_{k=0}(\Lambda,s)&=&1 
\eea 
The condition \eq{f1} ensures that the IR region is suppressed.
Notice that the limits $k\to 0$ and $s\to\infty$ do not commute, since
$\lim_{s\to\infty}f_{k= 0}(\Lambda,s)=1$. The condition \eq{f3}
implies that all traces regularised with $f_k$ vanish at this point.
Thus, at one-loop, we have a trivial initial condition
$\Gamma_{k=\Lambda}=S_{\rm cl}$.  Finally, the condition \eq{f4}
ensures that the proper-time regularisation reduces to the usual
Schwinger proper time regularisation for $k=0$. This implies that
$\Gamma_{k=0}[\phi]|_{\rm 1-loop}= \Gamma^{\rm 1-loop}[\phi;\Lambda]$.
Beyond one-loop, $\Gamma_k$ is not given as a closed expression but
only by the integrated flow and the initial condition
$\Gamma_\Lambda$. Thus \Es{f3} and \eq{f4} do not imply
$\Gamma_{k=\Lambda}=S_{\rm cl}$ and $\Gamma_{k=0}[\phi]=\Gamma[\phi]$,
the full effective action, as opposed to the ERG case.  We introduce
the dimensionless function $f_{\rm PT}(x)$ as
\beq\label{f}
f_k(\Lambda,s)=f_{\rm PT}(s\Lambda^2)-f_{\rm PT}(sk^2)\,  
\eeq 
for later convenience. It obeys $\partial_t f_k(\Lambda,s) =
-\partial_t f_{\rm PT}(sk^2)$ and $\partial_t f_{\rm PT}(x)=2xf_{\rm
  PT}'(x)$. In this parametrisation, only the function $f_{\rm
  PT}(sk^2)$ appears in the flow \eq{PTRG}.  The conditions \eq{f1} --
\eq{f4} translate into $f_{\rm PT}(x\to \infty)=1$ and $f_{\rm
  PT}(x\to 0)=0$.  \step

The flow \eq{PTRG} describes, at least, a resummation of a subset of
perturbative diagrams. Let us denote with $\Gamma_{\rm PT}$ the
solution for $k\to 0$ of the flow \eq{PTRG}. Since a derivation from
first principles is missing, it is {\it not} known how $\Gamma_{\rm
  PT}$ is related to the physical theory, {\it e.g.}~the full quantum
effective action $\Gamma$. Furthermore, $\Gamma_{\rm PT}$ may depend
on the specific regulator chosen.  These facts are a severe limitation
of the PTRG formalism; they restrict the predictive power, because no
control over the link of $\Gamma_{\rm PT}$ to the physical theory is
yet available.\step

In order to provide some insight into these issues we investigate the
relation of PTRG to ERG. Within the ERG approach it is known by
construction that the limit $\lim_{k\to 0}\Gamma_k=\Gamma_{\rm ERG}$
of the ERG flow coincides with the full quantum effective action
$\Gamma_{\rm ERG}\equiv\Gamma$. We exploit this piece of information
to provide a better understanding of the physical content of the PTRG.
\\[3ex]

\noindent
{\bf 4. Derivative expansion}\\[-1ex]

From now on, we restrict ourselves to the study of \Es{ERG} and
\eq{PTRG} within the leading order of the derivative expansion. The
derivative expansion is the most commonly used systematic
approximation scheme. It is most useful for theories which, except for
possible modifications due to anomalous dimensions, retain a standard
kinetic term in the IR limit. The derivative expansion has been used
very successfully for the computation of critical exponents or
equations of states for scalar theories.\step

For later use, we consider an $O(N)$ symmetric scalar field theory in
$d$ dimensions and to leading order in the derivative expansion. The
model has an effective action
\beq\label{AnsatzGamma}
\Gamma_k=\int d^dx \left[U_k(\bar\rho) 
             + \s012\partial_\mu \phi^a\partial_\mu \phi_a
             +{\cal O}(\partial^4)
\right]
\eeq
and $\bar\rho=\s012 \phi^a\phi_a$. We introduce dimensionless variables,
\bea
u(\rho)&=&U_k(\bar\rho)\, k^{-d}\\
\rho&=&\bar\rho\, k^{2-d}\,.
\eea
Inserting the Ansatz \eq{AnsatzGamma} into either \Eq{ERG} or
\Eq{PTRG} leads to
\beq\label{FlowPotential} 
\partial_t u+d u -(d-2) \rho u' 
=  2v_d(N-1)\, \ell (u') 
  +2v_d\,      \ell (u'+2\rho u'')\,, 
\eeq 
and $v_d^{-1}=2^{d+1}\pi^{d/2}\Gamma(\s0d2)$. All information
regarding the regulator function is contained in the function
$\ell(\omega)$. For the ERG case, it is given by
\beq
\label{ThresholdDef}
\ell_{\rm ERG}(\omega)
= \s012\int_0^\infty dy y^{\s0d2}\,\0{\partial_t r(y)}{y(1+r)+w}\,.
\eeq
Here, we have used \Eq{r}, and $y=q^2/k^2$ is the dimensionless
momentum variable, and $\partial_t r(y)=-2y r'(y)$. For the PTRG case,
$\ell(\omega)$ is given by
\beq
\label{ProperThreshold}
\ell_{\rm PT}(\omega)= \s012 \Gamma(\s0d2)\int_0^\infty
\0{dx}{x^{1+d/2}} 
\left[ \partial_t f_{\rm PT}(x)\right] \exp (-x \omega)\,,
\eeq
the PTRG counterpart of $\ell_{\rm ERG} (\omega)$. Here, we have used
\Eq{f}, and $x\equiv s k^2$ is the dimensionless integration variable
of the proper-time integral. Notice that the leading order flow
\eq{FlowPotential} is structurally the same for ERG and PTRG. This
suggests that the formalisms can be mapped onto each other to leading
order in the derivative expansion.
\\[3ex]

\noindent
{\bf 5. From ERG to PTRG}\\[-1ex]

Now we establish the map from ERG $\to$ PTRG to leading order in the
derivative expansion. Specifically, for every given regulator function
$R_k(q^2)$, we provide a corresponding function $f_k(s)$. The
construction is the following: we assume that a regulator $R_k(q^2)$
has been given. Therefore, the function $\ell_{\rm ERG}(\omega)$ is
known, either $(i)$ explicitly as an analytic function of $\omega$, or
$(ii)$ as an expansion in powers of $\s01{1+\omega}$, or $(iii)$ as an
expansion in powers of $\omega$.  Equating the function $\ell_{\rm
  ERG}(\omega)$ with $\ell_{\rm PT}(\omega)$ provides the map from the
momentum regulator $R_k(q^2)$ to the proper-time regulator
$f_k(\Lambda,s)$. \step

$(i)$ We begin with the simplest case, where $\ell_{\rm ERG}(\omega)$
is given as an explicit analytic function of $\omega$. In this case 
one can read off from 
\eq{ProperThreshold} that $\ell_{\rm ERG}(\omega)$ is the Laplace
transform of the function
\beq
g(x)=\Gamma(\s0d2)\, x^{-d/2}\, f_{\rm PT}'(x)\,.
\eeq 
Then it suffices to perform the inverse Laplace transformation in
order to obtain the corresponding $f_{\rm PT}'(x)$ and hence $f_{\rm
  PT}(x)$. Let us give three examples. We first consider the sharp
cut-off $R_{\rm sharp}=\lim_{c\to\infty} c\Theta(k^2-q^2)$. It leads
to the threshold function $\ell_{\rm sharp}(\omega)=-\ln (1+\omega)$.
Considering $\partial_w \ell_{\rm sharp}(\omega)$ (because the Laplace
transform of $\ln (1+\omega)$ does not exist) gives $g_{\rm
  sharp}(x)=-x^{-1} e^{-x}$ and the differential equation
\beq\label{SolutionS} 
\partial_t f_{\rm sharp}(x) =-\02{\Gamma(\s0d2)}\, x^{d/2}\, e^{-x}\,,  
\eeq 
in agreement with the corresponding result given in
Ref.~\cite{Liao:1996fp}. The second example concerns the power-like
regulator $R_{\rm power}(q^2)=q^2(k^2/q^2)^b$ for $b=2$ in $d=3$
dimensions \cite{Morris:1998xj}. It leads to $\ell_{\rm
  power}(\omega)=2\pi/\sqrt{2+\omega}$. The corresponding PT regulator
(for $d=3$) is found as
\beq
\partial_t f_{\rm power}(x)=8 x^2\,e^{-2x}\,.
\eeq
The third example is the optimised regulator $R_{\rm
  opt}(q^2)=(k^2-q^2)\Theta(k^2-q^2)$ \cite{Litim:2001up}. It leads to
$\ell_{\rm opt}(\omega)=\s02d{(1+\omega)^{-1}}$.  We find ${g_{\rm
    opt}(x)}=\s02d e^{-x}$ and therefore
\beq\label{Solution1} 
\partial_t f_{\rm opt}(x)
=\s04d\01{\Gamma(\s0d2)}\, x^{\s0d2+1}\, e^{-x} 
\eeq 
in agreement with the result given in Ref.~\cite{Litim:2001up}. \step

$(ii)$ Let us assume that $\ell_{\rm ERG}(\omega)$ is given
as the particular series
\beq\label{B-Series} 
\ell_{\rm ERG}(\omega)
= \s02d\0{1}{1+\omega}\sum_{n=0}^\infty b_n 
  \left(\0{-1}{1+\omega}\right)^n
\eeq 
with the expansion coefficients 
\beq
\label{bn}
b_n=\int^\infty_0 dy y^{d/2+1}{[-r'(y)]}{[y(1+r)-1]^n}\,.
\eeq
Such series exist for all ERG regulators that decay more than
power-like for large momenta and have a mass-like limit for small
momenta \cite{Litim:2001up}. In particular it follows from
\Eq{ThresholdDef} or \Eq{bn} that $b_0\neq 0$.  The basic set of
functions are the monomials $(1+\omega)^{-n}$ for which their Laplace
transform is known; hence
\beq
\label{ii} 
g(x)=\s02d \sum^\infty_{n=0} \0{b_n }{n!} (-x)^{n}\, e^{-x}\,.  
\eeq 
and 
\beq\label{f'-generic}
\partial_t f_{\rm PT}(x)
=\s04d \01{\Gamma(\s0d2)}\sum^\infty_{n=0} \0{b_n }{n!}
(-)^n x^{n+\s0d2+1}\, e^{-x}\,.
\eeq 
Finally, this gives $f_{\rm PT}(x)$ as an alternating sum 
\beq
\label{f-generic}
f_{\rm PT}(x)= 
\s04d \01{\Gamma(\s0d2)} 
\sum^\infty_{n=0} \0{b_n }{n!}
(-)^n\Gamma(n+\s0d2+2,x)  \quad {\rm with} \quad  b_0\neq 0
\eeq 
where $\Gamma(a,x)=\int_0^xdt\,t^{a-1}e^{-t}$ is the incomplete
$\Gamma$-function. Notice that \Eq{f'-generic} reduces to
\Eq{Solution1} for $b_0=1$ and all other $b_n=0$.\step

$(iii)$ The third case concerns those ERG regulators for which the
functions $\ell_{\rm ERG}(\omega)$ are not expandable as in
\Eq{B-Series}. A well-known example is given by the sharp cut-off, or
power-like regulators $R_k\sim q^2(k^2/q^2)^b$. However, the expansion
\beq\label{A-Series}
\ell_{\rm ERG}(\omega)= \s02d\sum_{n=0}^\infty a_n (-\omega)^n 
\eeq 
with the expansion coefficients 
\beq
\label{an}
a_n=\int^\infty_0 dy y^{d/2+1}\0{-r'(y)}{[y(1+r)]^n}\,.
\eeq
{\it always} exists for arbitrary regulators \cite{Litim:2000ci}.  In
this case the Laplace transform is of no help because the space of
functions spanned by $\{\omega^n\}$ is not Laplace transformable.
Still, we can compute all moments of the function $g(x)$, which are
given by
\beq\label{iii}
a_n=\0d2\0{1}{n!\,\Gamma(\s0d2)}\int_0^\infty dx x^n g(x) 
\eeq 
The reconstruction problem ({\i.e.~finding $g(x)$ and hence $f_{\rm
PT}(x)$ from the set $\{a_n\}$) is very similar to the
reconstruction of structure functions within perturbative QCD. Let
us assume that we know the set $\{a_n\,n\le N\}$ numerically up to
an order $N$. This implies that {\it infinitely} many functions
$g(x)$ can be found which all have the first $n$ moments \eq{iii}.
The map can be made unique only under additional assumptions
regarding the small-$x$ and the large-$x$ behaviour of $g(x)$,
provided by the large-$n$ behaviour of $a_n$. Here, we make the {\it
ad hoc} assumption that $g(x)=P_N(x)x^\alpha\exp(-x)$ with $P_N$ a
polynomial of order $N$. Then all coefficients of $P_N=\sum_{m=0}^N
\s01{m!}p_m x^m$ are determined as ${\bf p}={\bf M}^{-1}{\bf a}$,
with the numerical $(N+1)\times(N+1)$ matrix ${\bf M}$ given by
\beq\label{M} 
({\bf M})_{nm}=\0d2\0{\Gamma(m+n+\alpha+1)}{\Gamma(\s0d2)\,n!\,m!}\,.
\eeq 
The quality of the result would then depend on the assumptions
concerning the asymptotic behaviour (like the free parameter
$\alpha$). From the explicit solutions \eq{SolutionS} and
\eq{Solution1} we deduce that $\alpha=-1$ for the sharp cut-off, and
$\alpha=0$ for the optimal regulator. Deriving the correct value for
$\alpha$ from the large-$n$ behaviour of $a_n$ seems to be the most
difficult part of the reconstruction problem. Notice also that the
expansion \eq{B-Series} is much more powerful than the expansion
\eq{A-Series}, because the corresponding radius of convergence is
larger \cite{Litim:2001up}.
\\[3ex]
 
\noindent
{\bf 6. From PTRG to ERG}\\[-1ex]

Here, we show that the inverse map from PTRG to ERG does not exist in
general, not even to leading order in the derivative expansion. We
will not discuss the specific conditions required for $f_k(s)$ such
that this map may exist. Consider the function
\beq
\label{counter} 
f_{\rm PT}(x;m)= \0{\Gamma(m,x)}{\Gamma(m)} \, 
\eeq
with 
\beq\label{dt-counter} 
\partial_t f_{\rm PT}(x;m)= \0{2}{\Gamma(m)}
\,x^m\, e^{-x}\,.  
\eeq 
This set of functions is particularly important since it is the
standard set used for analytical considerations
\cite{Liao:1996fp,Liao:1997nm} or numerical applications
\cite{Papp:2000he,Schaefer:1999em,Bohr:2000gp,Bonanno:2001yp} of the
PTRG.  For $m\ge \s0d2$, this function satisfies the basic
requirements imposed on $f_k(\Lambda,s)$. It leads to the function
\beq
\label{ell-counter} 
\ell_{\rm PT}(\omega)
=\0{\Gamma(\s0d2)\Gamma(m-\s0d2)}{\Gamma(m)(1+\omega)^{m-d/2}} 
\eeq
with the asymptotic behaviour $\ell_{\rm PT}(\omega\to\infty)\sim
\omega^{-m+d/2}$ for arbitrary $m$. However, it follows from
\Eq{ThresholdDef}, that $\ell_{\rm ERG}( \omega\to\infty)> C\,
\omega^{-1}$, where $C>0$ depends on the regulator $R_k$. This is
easily deduced from \Eq{ThresholdDef}, if $r(y)$ is a monotonously
decreasing function in $y$ (and $\ln k$). It holds as well for
oscillating regulators $r$, as long as $y(1+r)$ is strictly positive.
Hence, the asymptotic decay of $\ell_{\rm ERG}( \omega)$ is at most
$\sim\omega^{-1}$ or weaker. Therefore, it is impossible to find the
ERG analogue to \Eq{counter} once $m>\s0d2+1$. It is interesting to
note that the optimised regulator of Ref.~\cite{Litim:2001up}
corresponds precisely to the boundary case $m=\s0d2+1$. \step

Let us now consider $m>\s0d2 +1$. The scale derivative $\partial_t
f_{\rm PT}(x;m)$ as given in \Eq{dt-counter} decays for both $x\to 0$
and $x\to \infty$ and has its maximum at $x=m$. Thus the PTRG flow
with a regulator \eq{dt-counter} has an effective IR scale $k_{\rm
  eff}\propto k/m^{1/2}$, since $x$ (roughly) corresponds to
$k^2/q^2$. This has already been noted in Ref.~\cite{Bonanno:2001yp}.
It is understood that the factor $m^{1/2}$ only takes care of the
leading $m$-dependence relevant for the limit of large $m\gg \s0d2+1$.
Hence, the flow, with increasing $m$ and fixed $k$, is peaked at
increasingly small momenta with decreasing width.  The above
explanations make it clear why such a flow, for sufficiently large
$m$, cannot be mapped onto ERG flows at an IR scale $k$. Moreover the
integrated flow (starting at a fixed initial scale $\Lambda$ and going
to $k=0$) is {\it not} covering the whole momentum regime, but only
the interval $[\Lambda_{\rm eff},0]$ with $\Lambda_{\rm eff}\propto
\Lambda/m^{1/2}$. Hence, the initial effective action
$\Gamma_\Lambda$, for consistency, has to contain all quantum effects
originating from the momentum interval $[\infty,\Lambda_{\rm eff}]$.
For $m\to \infty$ at fixed initial scale $\Lambda$, this means that
the starting point is the full quantum effective action.\step

Based on the vanishing width of the regulator \eq{dt-counter} for
$m\to\infty$, it has been suggested that it may correspond to a sharp
cut-off limit, and that this limit may provide a sensible regulator
for both UV and IR modes \cite{Bonanno:2001yp}. However, a decreasing
-- and eventually vanishing -- width occurs, by definition, if the
effective cut-off scale $k_{\rm eff}$ is removed.  Again a comparison
with the ERG is helpful: {\it any} smooth regulator $R_k(q^2)$ gets a
vanishing width for $k\to 0$. Moreover the flow at $k=0$, which still
depends on the specific form chosen for $R_k$, is not unique.  The
same applies to PTRG flows. Hence, the flow at $m=\infty$ at a finite
scale $k$ is neither sharp, nor exact (see also
Ref.~\cite{Consistency}). It is a flow at vanishing effective IR scale
$k_{\rm eff}=0$.\step

However, it is more desirable to consider flows where the effective
initial scale $\Lambda_{\rm eff}$ and the effective infrared scale
$k_{\rm eff}$ are independent of the choice for the regulator. For the
class of regulators given by \Eq{counter}, this is achieved by
introducing the effective scales $k_{\rm eff}$ and $\Lambda_{\rm eff}$
according to $k= m^{1/2} k_{\rm eff}$, and hence $\Lambda=
m^{1/2}\Lambda_{\rm eff}$.  The corresponding effective action is
$\hat\Gamma_{k_{\rm eff}}=\Gamma_{m^{1/2}k_{\rm eff}}$.  As has been
argued before, it includes all quantum effects of momenta larger than
$k_{\rm eff}$.  Furthermore, we note that the width of the flow stays
finite for any $m$ and fixed effective scales. In order to confirm
this picture, we introduce $(\Lambda_{\rm eff},k_{\rm eff})$ as
described above, but denote them as $(\Lambda,k)$ for notational
simplicity. After these manipulations, and using \Es{PTRG},
\eq{counter} and \eq{dt-counter}, the flow for $\hat\Gamma_k$ becomes
\begin{eqnarray}\label{FlowScale}
\partial_{t}\hat\Gamma_{k}=
\int_0^\infty d s
\,\0{(m\, s k^2)^m\ e^{-m\, s k^2}}{\Gamma(m)}
\,\Tr\, \exp\left(-s\hat\Gamma_{k}^{(2)}\right).
\end{eqnarray}
The prefactor in front of the trace is $\s012\partial_t f_{\rm PT}(
m\,x;m)$ with $x=sk^2$, and has the simple limit $\lim_{m\to
  \infty}\s012 f_{\rm PT}(m\,x;m)=\delta(x-1)$.  This follows from the
asymptotic behaviour of the $\Gamma$ function $\Gamma(m\to\infty)\to
\sqrt{2\pi}\, m^{1/2}\, m^m\, e^{-m}$.  Thus, for $m\to\infty$ we
arrive at
\beq\label{FlowExp}
\partial_{t}\hat\Gamma_{k}=
\Tr\, \exp\left(-\hat\Gamma_{k}^{(2)}/k^2\right).
\eeq 
\Eq{FlowExp} is the closed form of the integral equation \eq{PTRG} for
$m=\infty$ at the effective cut-off scale $k_{\rm eff}$. No
approximation to the full flow related to $\partial_t f_{\rm
  PT}(x;m=\infty)$ is made. Note also that we could have started with
the relation $k\to \alpha\,m^{1/2}\,k$ leading to $1/(\alpha^2\,k^2)$
in the exponent in \eq{FlowExp}. This amounts to a redefinition of
$\hat\Gamma_k$ and displays some lack of information about the initial
effective action $\hat\Gamma_\Lambda$.  \step

Now we are in the position to compare ERG flows for general regulators
$R_k(q^2)$, and PTRG flows based on $f_{\rm PT}(x;m)$ as defined in
\Eq{dt-counter} at a fixed cut-off scale relevant for both flows.
After a fixed effective scale is taken into account, the functions
$\ell_{\rm PT}(\omega)$ effectively depend on $\omega/m$. They still
decay faster than $\omega^{-1}$ for $m>\s0d2+1$. This remains so even
for $m\to\infty$, where $\ell_{\rm
  PT}(\omega)=\Gamma(\s0d2)\,e^{-\omega}$. Hence also for $m=\infty$
the function $\ell_{\rm PT}$ does not meet the decay condition
$\ell_{\rm ERG}(w\to \infty)> C\omega^{-1}$ necessary for having an
ERG analogue in the approximation studied here.
\\[3ex]

\noindent
{\bf 7. Comparison of critical exponents}\\[-1ex]

We now turn to the computation of universal critical exponents of
$O(N)$-symmetric scalar theories in $3d$ to leading order in the
derivative expansion, based on the $m$-dependent flows
\eq{ell-counter}. The range $\s0d2\le m\le \s0d2+1$ corresponds to
approximations of exact flows, and the corresponding results can be
taken as predictions. In turn, it is not known how the flows for the
parameter range $\s0d2+1 < m\le \infty$ relate to approximations of
exact flows.  Therefore, the corresponding results have to be taken
with some reservations.  \step

\begin{figure}[t]
\begin{center}
\unitlength0.001\hsize
\begin{picture}(500,580)
\put(400,500){{\huge $ \nu$}}
\put(150,530){{$N=4$}}
\put(210,440){{$3$}}
\put(210,360){{$2$}}
\put(210,280){{$1$}}
\put(150,190){{$N=0$}}
\put(280,80){\large $m$}
\put(475,120){\large $\infty$}
\psfig{file=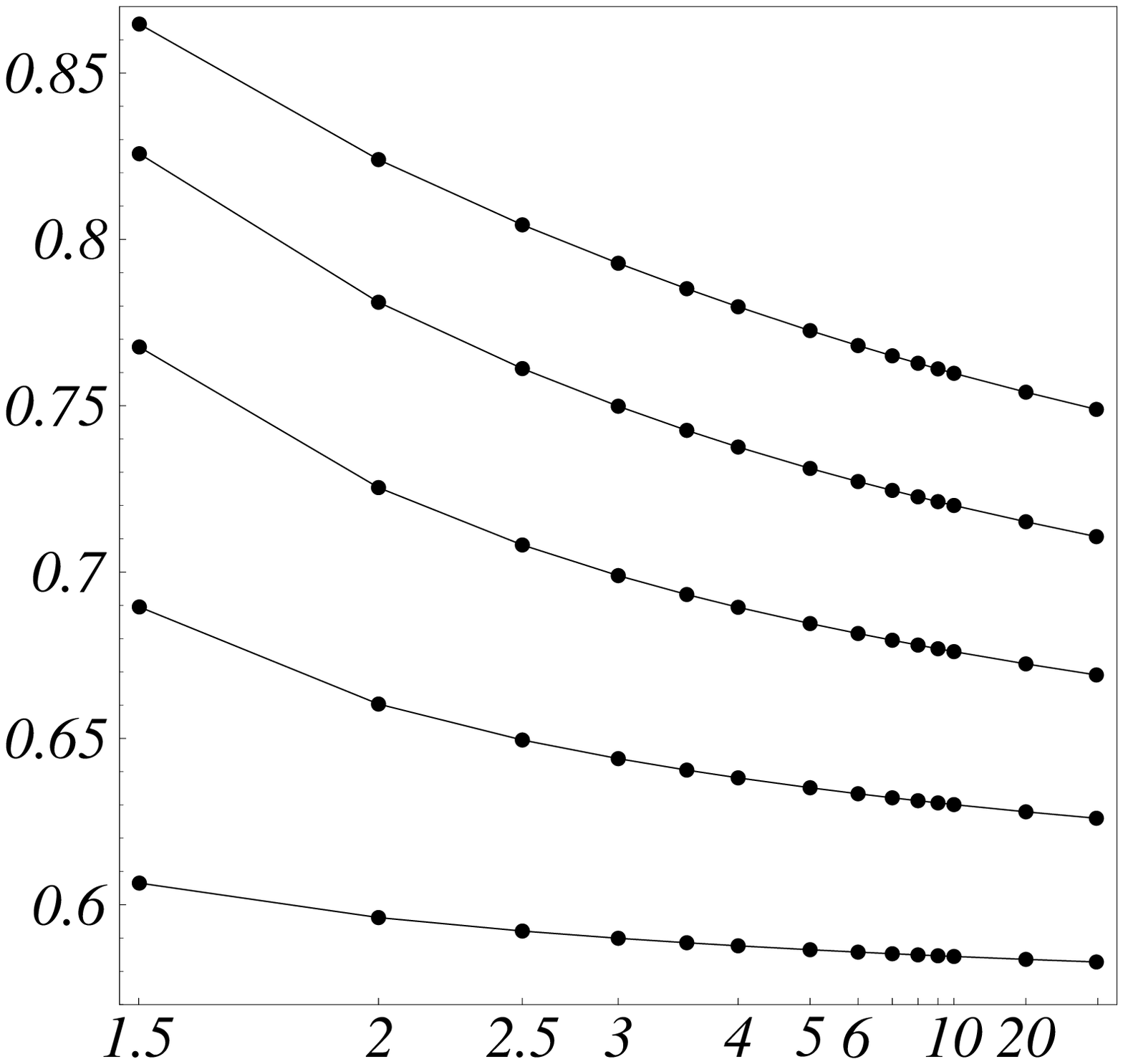,width=.5\hsize}
\end{picture}
\vskip-.5cm
\begin{minipage}{.92\hsize}
  {\bf Figure 1:} The dependence of $\nu$ on the parameter $m$. The
  $m$-axis has been squeezed as $m\to (m-\s032)/m$ for display
  purposes.
\end{minipage} 
\end{center}
\end{figure}

\begin{figure}[t]
\begin{center}
\unitlength0.001\hsize
\begin{picture}(500,580)
\put(350,500){{\huge $ \omega$}}
\put(100,530){{$N=4$}}
\put(110,450){{$3$}}
\put(110,360){{$2$}}
\put(110,280){{$1$}}
\put(110,220){{$0$}}
\put(280,80){\large $m$}
\put(475,120){\large $\infty$}
\psfig{file=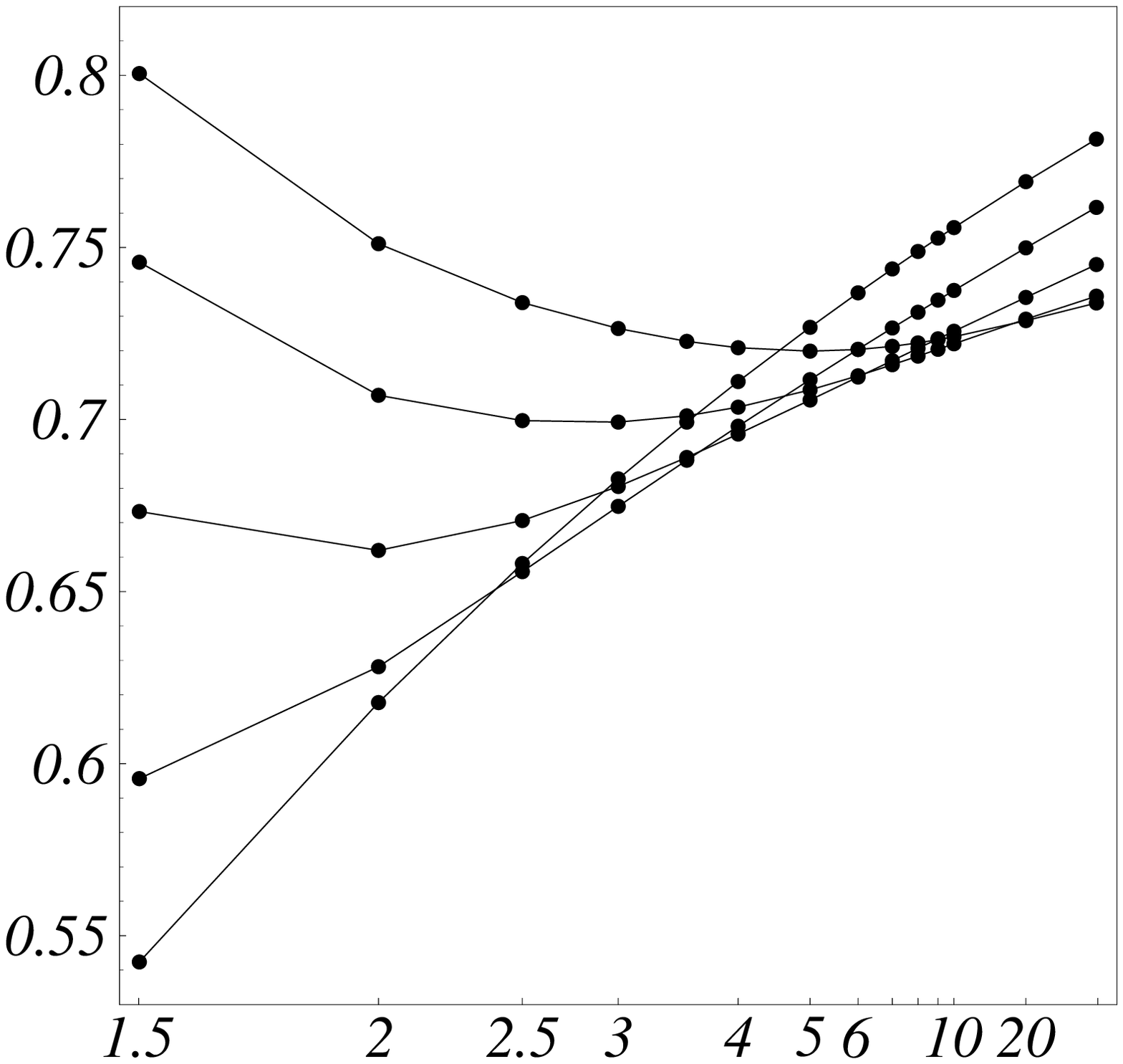,width=.5\hsize}
\end{picture}
\vskip-.5cm
\begin{minipage}{.6\hsize}
  {\bf Figure 2:} The dependence of $\omega$ on the parameter $m$. 
\end{minipage} 
\end{center}
\end{figure}

The physically most interesting cases are the universality classes
$N=0$ (polymers), $N=1$ (Ising model), $N=2$ (XY-model), $N=3$
(Heisenberg model) and $N=4$, which is expected to be relevant to the
thermal QCD phase transition with two light quark flavours. The
Wilson-Fisher fixed point corresponds to the non-trivial fixed point
solution $\partial_t u_\star=0$. The scaling solution $u_\star$ has
one unstable eigen-direction with eigenvalue $\lambda_0<0$. Its
negative inverse is given by the critical index $\nu=-1/\lambda_0$.
The smallest positive eigenvalue is denoted as $\omega$.  For our
computation of $\nu$ and $\omega$ we use the flow \eq{FlowPotential}
with the regulator \eq{counter} and $m\ge\s032$. For $d=3$, the flow
is given by
\beq\label{Flow3} 
\partial_t u+3 u - \rho u' =
\0{(N-1)\Gamma(m-\s032)}{8\pi^{3/2}\Gamma(m)(1+u')^{m-3/2}}
+\0{\Gamma(m-\s032)}{8\pi^{3/2}\Gamma(m)(1+u'+2\rho u'')^{m-3/2}}\,.
\eeq 
The case $m=\s032$ corresponds to the sharp cut-off
\cite{Wilson}, $m=2$ to the quartic regulator $R=k^4/q^2$
\cite{Morris:1998xj}, and $m=\s052$ to the optimal regulator $R_{\rm
  opt}$ \cite{Litim:2001up}.\step  

For $m\to\infty$ we can rely on \Eq{FlowExp} which, within the given
approximation, leads to
\beq\label{Flow3Infty} 
\partial_t u+3u -\rho u' =
(N-1)\exp(-u')+\exp(-u'-2\rho u'')\,,  
\eeq
where the variables $u,\rho$ are those of $\hat\Gamma_{k}$ and we have
redefined $u\to u/(8 \pi^{3/2})$ and $\rho\to\rho/(8 \pi^{3/2})$.  Of
course one also can obtain \Eq{Flow3Infty} from \Eq{Flow3}. This is
seen as follows. The dimensionless fixed point potential is non-convex
(convex) in those regions of field space where $u'$ or $u'+2\rho u''$
are negative (positive). Hence, increasing $m$ in \Eq{Flow3} has three
effects.  First, it leads to a parametrical suppression of the
right-hand side of \Eq{Flow3}, proportional to
$\Gamma(m-\s032)/\Gamma(m)$. Second, the contributions from the convex
part of the potential are strongly suppressed. Third, the
contributions from the non-convex region, which shrinks as
$\Gamma(m-\s012)/\Gamma(m)$, are strongly enhanced. Therefore, we
redefine the field variable in \Eq{Flow3} as $\rho\to
\rho\,\Gamma(m-\s012)/(8\pi^{3/2}\Gamma(m))$ and the potential as
$u\to u\,\Gamma(m-\s032)/(8\pi^{3/2}\Gamma(m))$. It is worth pointing
out that this redefinition of $u$ and $\rho$ precisely amount to the
redefinition $k\to m^{1/2} k$ within the derivation of \Eq{FlowExp}
and the finite redefinition of field variables as explained directly
below \Eq{Flow3Infty}.\step

We have computed the critical indices $\nu$ and $\omega$ from
\Eq{Flow3} for $N=0\cdots 4$ and for $m=\s032$ up to $m=10^6$. For the
case $m=\infty$ we use the flow \eq{Flow3Infty}. In practice, instead
of solving numerically the flow $\partial_t u$, we found it more
convenient to solve the flows for the $\rho$-derivative $\partial_t
u'$. The numerical results for $\nu$ and $\omega$ are given in Tab.~1
and Tab.~2, respectively.\footnote{For $m=\s032$ and $m=2$, and for
  $N=1\cdots 4$, we have checked our numerical code with earlier
  findings summarised in Ref.~\cite{Comellas:1997tf}. For $2\le m\le
  10$ and $N=1$, we have compared our results with those of
  Ref.~\cite{Bonanno:2001yp}.  Notice that our parameter $m$ is
  related to the one of Ref.~\cite{Bonanno:2001yp} by $m=m_{\rm
    BZ}+1$.}  Notice that $m=10^3$ is not yet sufficiently large to
determine the first four significant figures for $\nu$ and $\omega$
corresponding to $m=\infty$ for all $N$ considered. However, for
$m=10^6$, the critical indices obtained from \Eq{Flow3} agree to (at
least) four significant figures with the result obtained from
\Eq{Flow3Infty}.  \step

Let us first discuss our results for $\nu$ and $\omega$ as displayed
in Figs.~1 and 2 as functions of $m$. The critical index $\nu$ is
monotonously decreasing with $m$ for all values of $N$. For $\nu$, the
convergence to the asymptotic value at $m=\infty$ is slightly better
for smaller values of $N$. In contrast, we notice that the first
irrelevant eigenvalue $\omega$ is no longer a monotonous function of
$m$, because its $m$-derivative changes sign for $N=2,3$ and $4$.
From Tab.~1 and 2, we deduce that the limit for $m\to\infty$ is
approached smoothly. However, the variation with $m$, over the entire
scale, is roughly twice as big for $\omega$ as compared to $\nu$.
\step

The results depend on the unphysical parameter $m$.  The dependence of
physical observables on unphysical parameters is well-known from
perturbative QCD, e.g.~Ref.~\cite{Stevenson}. Here, it is a side
effect of unavoidable approximations \cite{Scheme}. It may seem, at
first sight, that an unphysical scheme dependence restrains the
predictive power. However, in the context of the ERG, it has been
shown that this is not the case
\cite{Litim:2000ci,Litim:2001up,Litim:2001fd}. Indeed, the convergence
of approximate solutions is partly controlled by the regulator.
Therefore, an adequate choice of the latter improves the convergence.
These findings suggested that the regulator $R_k$ can be used to
``optimise'' the physical content within a given approximation. For a
generic optimisation criterion, based on the regularised inverse
propagator, and which improves the convergence towards the physical
theory, we refer to Refs.~\cite{Litim:2000ci,Litim:2001up}. The result
for the optimised regulator Refs.~\cite{Litim:2001up,Litim:2001fd}
corresponds to $m_{\rm opt}=\s052$. As stated in
Ref.~\cite{Litim:2001fd}, we confirm that the critical exponents for
$m_{\rm opt}$ are indeed the smallest of all values for $\nu$ within
the range accessible to the ERG. \step

\begin{center}
\begin{tabular}{c|ccccc}
\hline\hline\\[-1ex]
$\quad{}       m                  \quad{}$&
$\quad{}  \nu_{\rm N=0}           \quad{}$&
$\quad{}  \nu_{\rm Ising}         \quad{}$&
$\quad{}  \nu_{\rm XY}            \quad{}$&
$\quad{}  \nu_{\rm Heisenberg}    \quad{}$&
$\quad{}  \nu_{\rm N=4}           \quad{}$
\\[.5ex] \hline\\[-1.5ex] 
$\s032$
&.6066
&.6895
&.7678
&.8259
&.8648
\\
2
&.5961
&.6604
&.7253
&.7811
&.8240
\\
$\s052$
&.5921
&.6496
&.7082
&.7611
&.8043
\\[.5ex] \hline\\[-1.5ex] 
10
&.5845
&.6301
&.6760
&.7199
&.7598         
\\
$10^2$
&.5829
&.6264
&.6696
&.7112        
&.7496
\\
$10^3$
&.5828
&.6260
&.6690
&.7104
&.7486
\\
$10^6$
&.5828
&.6260
&.6690
&.7103
&.7485
\\
$\infty $
&.5828
&.6260
&.6690
&.7103
&.7485
\\[1ex]\hline\hline
\end{tabular}
\end{center}

\begin{center}
\begin{minipage}{.95\hsize}
\vskip.3cm

{\label{Tab1}\small {\bf Table 1:} The critical exponent $\nu$ in
  $3d$ as a function of $m$ for different values of $N$. The case
  $m=\s032$ corresponds to the Wegner-Houghton equation, $m=2$ to an
  ERG flow with quartic regulator, and $m=\s052$ to the optimised ERG
  flow of Ref.~\cite{Litim:2001up}.  }

\end{minipage}
\end{center}

In the context of the PTRG, the values $m>\s052$ are also allowed.
However, the dependence on $m$ has a qualitatively different aspect:
the PTRG flow is not an exact one, which implies that the endpoint
$\Gamma_{\rm PT}$ depends on $m$ as well. Hence, the $m$-dependence of
$\nu$ and $\omega$ cannot be understood in the same way as within the
ERG; there, we took advantage of the fact that the endpoint of the ERG
flow is the full quantum effective action.  \step

Therefore, we employ a principal of minimum sensitivity (PMS)
condition \cite{Stevenson}, in order to single-out specific values for
$m$.\footnote{Within the PTRG approach, the scheme dependence has been
  addressed in Ref.~\cite{Papp:2000he} for $\s032\le m\le \s072$.
  Within the ERG formalism, a PMS condition has been used in
  Ref.~\cite{Liao:2000sh}. In Ref.~\cite{Litim:2001fd}, it has been
  explained why a PMS condition works for the ERG, and how it relates
  to the generic optimisation of Ref.~\cite{Litim:2000ci}.} Hence, we
will assume that the physical content of all $m$-dependent flows are
equivalent. Only then it is sensible to require that physical
observables should not depend on $m$, e.g.~$\partial \nu/\partial m
=0$ or $\partial\omega/\partial m =0$. From Tab.~1, we conclude that
$\partial \nu/\partial m <0$ for all $\s032<m<\infty$.  Hence, $\nu$
reaches its extrema at the boundary values $m=\s032$ and $m=\infty$.
For $\omega$, we notice that $\partial \omega/\partial m$ changes sign
for $N\ge 2$. Hence, for $N<2$ the boundary values $m=\s032$ and
$m=\infty$ are distinguished.  For $N\ge 2$ a true minimum at an
intermediate value of $m$ appears, and the boundary value at
$m=\infty$ (for $N=2$) or at $m=\s032$ (for $N=3,4$) corresponds to
the maximum. We conclude from the facts that a PMS condition does not
lead, for all observables, to a unique prescription for $m$. However,
the endpoints given by the sharp cut-off $m=\s032$, and by $m=\infty$,
are singled out because they represent (at least local) extrema for
all observables studied.  We emphasise that the optimised case $m_{\rm
  opt}=\s052$ -- which follows within the ERG formalism, and which is
closely linked to a PMS condition within the ERG approach
\cite{Litim:2001fd} -- does {\it not} follow from a PMS condition
within the PTRG approach. \step

\begin{center}
\begin{tabular}{c|ccccc}
\hline\hline\\[-1ex]
$\quad{}       m                      \quad{}$&
$\quad{}   \omega_{\rm N=0}           \quad{}$&
$\quad{}   \omega_{\rm Ising}         \quad{}$&
$\quad{}   \omega_{\rm XY}            \quad{}$&
$\quad{}   \omega_{\rm Heisenberg}    \quad{}$&
$\quad{}   \omega_{\rm N=4}           \quad{}$
\\[.5ex]   \hline\\[-1.5ex] 
$\s032$
&.5432
&.5952
&.6732
&.7458
&.8007
\\
2
&.6175
&.6286
&.6621
&.7068
&.7519
\\
$\s052$
&.6579
&.6557
&.6712
&.6998
&.7338
\\[.5ex] \hline\\[-1.5ex] 
10
&.7559
&.7376
&.7250
&.7200 
&.7231
\\
$10^2$
&.7794
&.7598
&.7437
&.7330
&.7290
\\
$10^3$
&.7816
&.7620
&.7455
&.7344
&.7297
\\
$10^6$
&.7819
&.7622
&.7457
&.7346
&.7298
\\
$\infty $
&.7819
&.7622
&.7457
&.7346
&.7298
\\[1ex]\hline\hline
\end{tabular}
\end{center}

\begin{center}
\begin{minipage}{.95\hsize}
\vskip.3cm

{\label{Tab2}\small {\bf Table 2:} The first irrelevant eigenvalue
  $\omega$ in $3d$ as a function of $m$ for different values of $N$.}

\end{minipage}
\end{center}

Finally, we compare the results of Tabs.~1 and 2 with those of other
methods (see Ref.~\cite{Pelissetto:2000ek} for a recent overview).
From all ERG flows, the results for $\nu$ with $m=\s052$ are closest
to those obtained by other methods. This confirms the conjecture that
optimised ERG flows have an improved derivative expansion
\cite{Litim:2001fd}.  For PTRG flows, it is intriguing to realise that
the values obtained for $m=\infty$ are in even better agreement with
both experimental values and those obtained by other techniques.
\\[3ex]

\noindent 
{\bf 8. Conclusions}\\[-1ex]

We investigated the predictive power of the PTRG by providing its link
to the ERG at leading order in the derivative expansion. We found that
the space of PTRG regulator functions $f_{\rm PT}$ is larger than the
space of ERG regulators $R_k$. Given the heuristic derivation of the
PTRG flow, there is no additional criteria available which would allow
to discard a specific subset of PTRG regulators. The ERG regulators,
however, have a simple physical interpretation. With help of the map
from ERG to PTRG -- even though only in the approximation discussed
here -- we can identify the subset of PTRG flows which are directly
related to momentum shell integrations in the Wilsonian sense. This
map cannot be extended to the full flow. \step

PTRG flows at a given IR scale $k$ for $m>\s0d2+1$ cannot be mapped to
ERG flows at $k$ even at leading order in the derivative expansion.
Still, they have a simple interpretation in terms of flows at a lower
effective IR scale $k_{\rm eff}\propto k/ m^{1/2}$.  Increasing $m$
changes the shape of the momentum cut-off {\it and} the initial
momentum scale $\Lambda_{\rm eff}\propto \Lambda/ m^{1/2}$, and hence
the initial action.  In terms of the fixed effective scales
$\Lambda_{\rm eff}$ and $k_{\rm eff}$, the PTRG flow \eq{FlowScale}
corresponds to a momentum shell integration at the scale $k_{\rm eff}$
starting at $\Lambda_{\rm eff}$.  It is precisely this picture which
stands behind the flow \Eq{FlowExp}, and the results given for
$m=\infty$. However, we still cannot map these flows to ERG flows at
the scale $k_{\rm eff}$.  Thus it remains unclear what kind of
approximation to flows of the full theory they correspond to.\step

As an application, we have computed critical exponents for $3d$
$O(N)$-symmetric scalar theories. Within the ERG, we have seen that
the optimisation of Ref.~\cite{Litim:2001up} indeed leads to improved
results, which correspond to the choice $m=\s0d2+1$. Within the PTRG,
we found very good results in the limit $m\to \infty$. While our above
findings suggest that this limit is sensible, it remains to be seen
how approximations to PTRG flows for $m>\s0d2+1$ are related to
systematic approximations of the physical theory, before these results
can be considered as predictions. \step

Our findings made clear that the precise structure of the inherit
approximation of the PTRG still have to be investigated more deeply in
order to use this approach in a well-controlled way. These inherit
approximations strongly depend on the regulator $f_k$, which makes the
task even more difficult. A possible avenue to escape from this
problem lies in a structural comparison of ERG and PTRG.  If one can
cast both equations into {\it similar} closed expressions, a full
discussion of similarities and differences is possible. Then one could
hope to derive quality checks for the PTRG in a closed form.  These
issues will be discussed in \cite{Consistency}. \\[3ex]

\noindent 
{\bf Acknowledgements}\\
DFL thanks the Institute for Theoretical Physics III, University of
Erlangen, and JMP thanks CERN for hospitality and financial support.
The work of DFL has been supported by the European Community through
the Marie-Curie fellowship HPMF-CT-1999-00404.\\[2ex]

\noindent 
{\bf Note added}\\
After this work was completed, the preprint \cite{Mazza:2001bp}
appeared, which also treats the limit $m\to\infty$.



\begin{thebibliography}{99}

\bibitem{Wilson}
F.J.\,Wegner and A.\,Houghton, Phys.~Rev.{\bf A8} (1973) 401; \\
K.G.\,Wilson and I.G.\,Kogut, Phys.~Rep~{\bf 12} (1974) {75}.

\bibitem{Polchinski}
J.\,Polchinski, Nucl.~Phys.~{\bf B231} (1984) 269.

\bibitem{CW}
C.\,Wetterich, Phys.~Lett.~{\bf B301} (1993) 90.

\bibitem{Ellwanger:1994mw}
U.~Ellwanger,
Z.\ Phys.\ C {\bf 62} (1994) 503
[hep-ph/9308260].

\bibitem{Morris:1994qb}
T.~R.~Morris,
Int.\ J.\ Mod.\ Phys.\ {\bf A9} (1994) 2411
[hep-ph/9308265].

\bibitem{Bagnuls:2000ae}
C.~Bagnuls and C.~Bervillier,
hep-th/0002034,\\
J.~Berges, N.~Tetradis and C.~Wetterich,
hep-ph/0005122.

\bibitem{Litim:1998nf}
D.~F.~Litim and J.~M.~Pawlowski,
{\it On gauge invariant Wilsonian flows},
hep-th/9901063.

\bibitem{Scheme}
R.~D.~Ball, P.~E.~Haagensen, J.~I.~Latorre and E.~Moreno,
Phys.\ Lett.\  {\bf B347} (1995) 80,\\
D.~F.~Litim,
Phys.\ Lett.\  {\bf B393} (1997) 103
[hep-th/9609040],\\
F.~Freire and D.~F.~Litim, 
hep-ph/0002153, to appear in Phys.~Rev.~{\bf D},\\
J.~Sumi, W.~Souma, K.~Aoki, H.~Terao and K.~Morikawa,
hep-th/0002231,\\
J.~I.~Latorre and T.~R.~Morris,
JHEP {\bf 0011} (2000) 004
[hep-th/0008123].


\bibitem{Litim:2000ci}
D.~F.~Litim,
Phys.\ Lett.\  {\bf B486} (2000) 92
[hep-th/0005245].

\bibitem{Litim:2001up}
D.~F.~Litim,
hep-th/0103195.

\bibitem{Litim:2001fd}
D.~F.~Litim,
hep-th/0104221, and in preparation.

\bibitem{Liao:1996fp}
S.~Liao,
Phys.\ Rev.\  {\bf D53} (1996) 2020
[hep-th/9501124].

\bibitem{Liao:1997nm}
S.~Liao,
Phys.\ Rev.\ D {\bf 56} (1997) 5008 
[hep-th/9511046].





\bibitem{Fock:1937dy}
V.~Fock,
Phys.\ Z.\ Sowjetunion {\bf 12} (1937) 404.


\bibitem{Schwinger:1951nm}
J.~Schwinger,
Phys.\ Rev.\  {\bf 82} (1951) 664.





\bibitem{Oleszczuk:1994st}
M.~Oleszczuk,
Z.\ Phys.\  {\bf C64} (1994) 533.

\bibitem{Consistency}
D.~F.~Litim and J.~M.~Pawlowski, in preparation.  

\bibitem{Stevenson}
P.~M.~Stevenson,
Phys.\ Rev.\  {\bf D23} (1981) 2916.


\bibitem{Liao:2000sh}
S.~Liao, J.~Polonyi and M.~Strickland,
Nucl.\ Phys.\ B {\bf 567} (2000) 493
[hep-th/9905206].

\bibitem{Papp:2000he}
G.~Papp, B.~J.~Sch\"afer, H.~J.~Pirner and J.~Wambach,
Phys.\ Rev.\ D {\bf 61} (2000) 096002
[hep-ph/9909246].




\bibitem{Floreanini:1995aj}
R.~Floreanini and R.~Percacci,
Phys.\ Lett.\ B {\bf 356} (1995) 205
[hep-th/9505172].


\bibitem{Morris:1998xj}
T.~R.~Morris and M.~D.~Turner,
Nucl.\ Phys.\ B {\bf 509} (1998) 637
[hep-th/9704202].

\bibitem{Comellas:1997tf}
J.~Comellas and A.~Travesset,
Nucl.\ Phys.\ B {\bf 498} (1997) 539
[hep-th/9701028].

\bibitem{Schaefer:1999em}
B.~Sch\"afer and H.~Pirner,
Nucl.\ Phys.\ A {\bf 660} (1999) 439
[nucl-th/9903003].

\bibitem{Bohr:2000gp}
O.~Bohr, B.~J.~Sch\"afer and J.~Wambach,
hep-ph/0007098.

\bibitem{Bonanno:2001yp}
A.~Bonanno and D.~Zappal\`a,
Phys.\ Lett.\ B {\bf 504} (2001) 181
[hep-th/0010095].

\bibitem{Pelissetto:2000ek}
A.~Pelissetto and E.~Vicari,
cond-mat/0012164.

\bibitem{Mazza:2001bp}
M.~Mazza and D.~Zappal\`a,
hep-th/0106230.
\end{thebibliography}
\end{document}